\documentclass[preprint,12pt]{elsarticle}
\usepackage{amssymb}
\usepackage{amsmath}
\journal{Subatomic Particles and Cosmology}
\begin{document}
\begin{frontmatter}
\title{Exploring the effects of $\alpha$-clustered structure of \textsuperscript{16}O nuclei in anisotropic flow fluctuations in \textsuperscript{16}O-\textsuperscript{16}O collisions at the LHC within a CGC+Hydro framework}
\author{Suraj Prasad\textsuperscript{a,b,\ref{note0}}, Neelkamal Mallick\textsuperscript{a,b,c},
Raghunath Sahoo\textsuperscript{a}, and \\Gergely Gábor Barnaföldi\textsuperscript{b}} 
\affiliation{organization={Department of Physics, Indian Institute of Technology Indore}, \\ 
            addressline={Simrol}, 
            city={Indore},
            postcode={453552}, 
            state={Madhya Pradesh},
            country={India}}
\affiliation{organization={HUN-REN Wigner Research Center for Physics}, \\
            addressline={29-33~Konkoly-Thege~Miklós~Str.}, 
            city={Budapest},
            postcode={H-1121},
            country={Hungary}}
\affiliation{organization={Department of Physics, University of Jyväskylä}, \\q
            addressline={P.O.~Box~35}, 
            city={Jyväskylä},
            postcode={FI-40014},
            country={Finland}}
\begin{abstract}
In this paper, we explore the effects of the presence of clustered nuclear structure of \textsuperscript{16}O in the final state elliptic flow fluctuations through \textsuperscript{16}O-\textsuperscript{16}O collisions at $\sqrt{s_{\rm NN}}=7$~TeV within a hybrid model, IPGlasma+MUSIC+iSS+ UrQMD. We compare the results of elliptic flow fluctuations using $\alpha$-clustered nuclear structure to the Woods-Saxon nuclear profile having no clustered structure. We observe a significant difference in the elliptic flow fluctuations, which arise due to the consideration of a clustered nuclear structure of~\textsuperscript{16}O.
\end{abstract}
\end{frontmatter}
\footnotetext[1]{Email: Suraj.Prasad@cern.ch\label{note0}}

\section{Introduction, methodology and event generation}
\label{sec1}
The observation of quark-gluon plasma (QGP)-like signatures in small systems is topical and piqued the interest of the collider physics community~\cite{ALICE:2016fzo}. Both the LHC and RHIC are colliding the \textsuperscript{16}O to explore small collision systems, which bridges the multiplicity gap between pp, p-Pb and Pb-Pb collisions~\cite{Brewer:2021kiv}. In addition, \textsuperscript{16}O possesses the clusters of $\alpha$-particles, which arrange themselves at the corners of a regular tetrahedron. Although the study of nuclear density profile is a matter of low-energy nuclear physics, in this paper, we explore the effect of clustered nuclear structure of \textsuperscript{16}O in elliptic flow fluctuations. 
\begin{figure}
    \centering
    \includegraphics[scale=0.26]{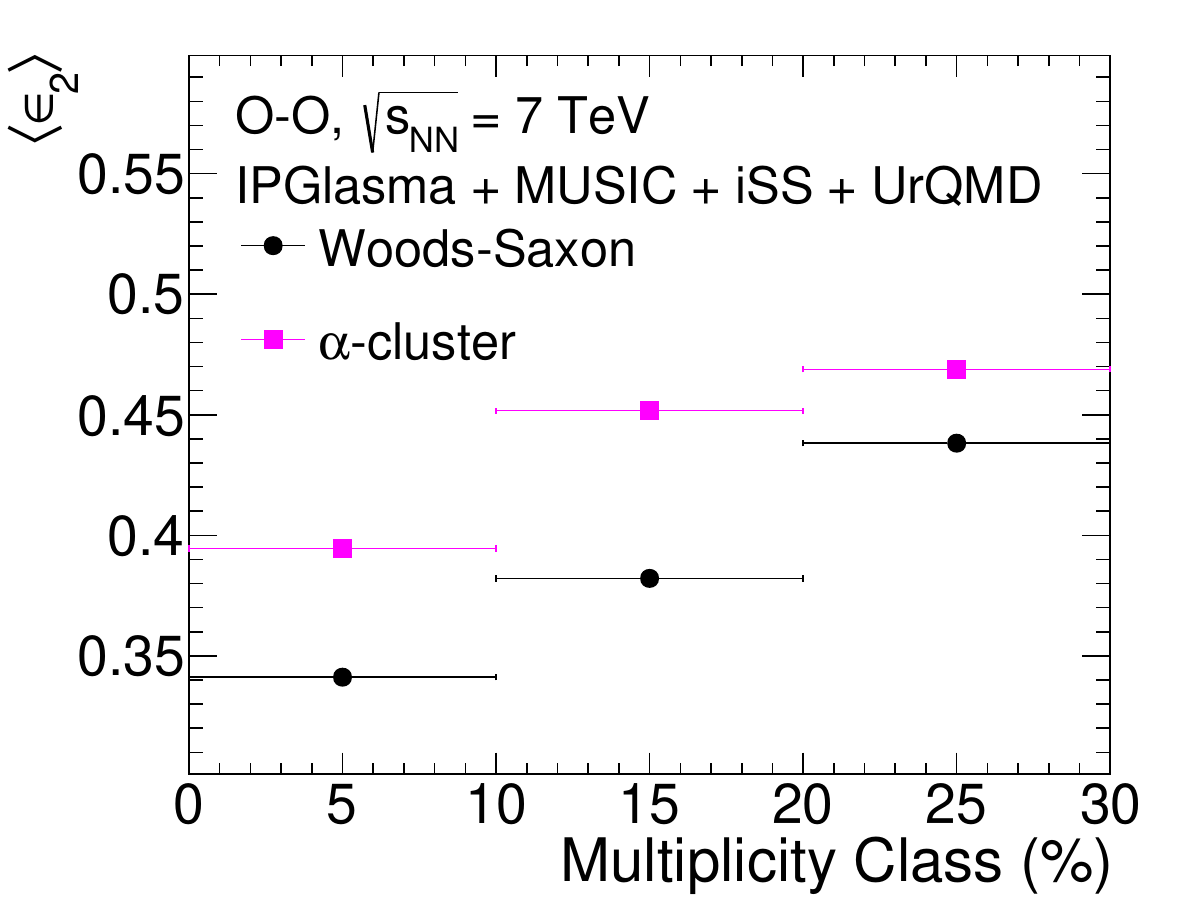}
    \includegraphics[scale=0.26]{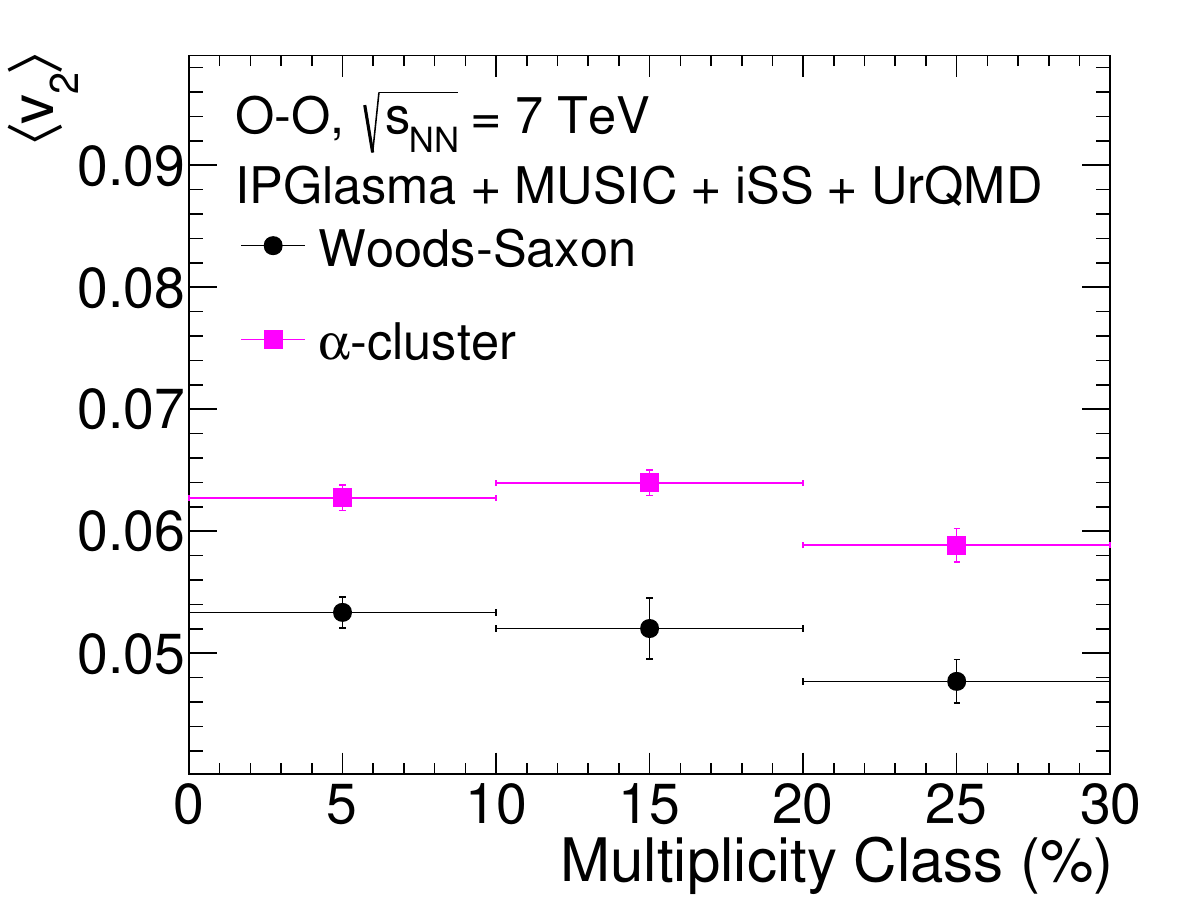}
    \includegraphics[scale=0.26]{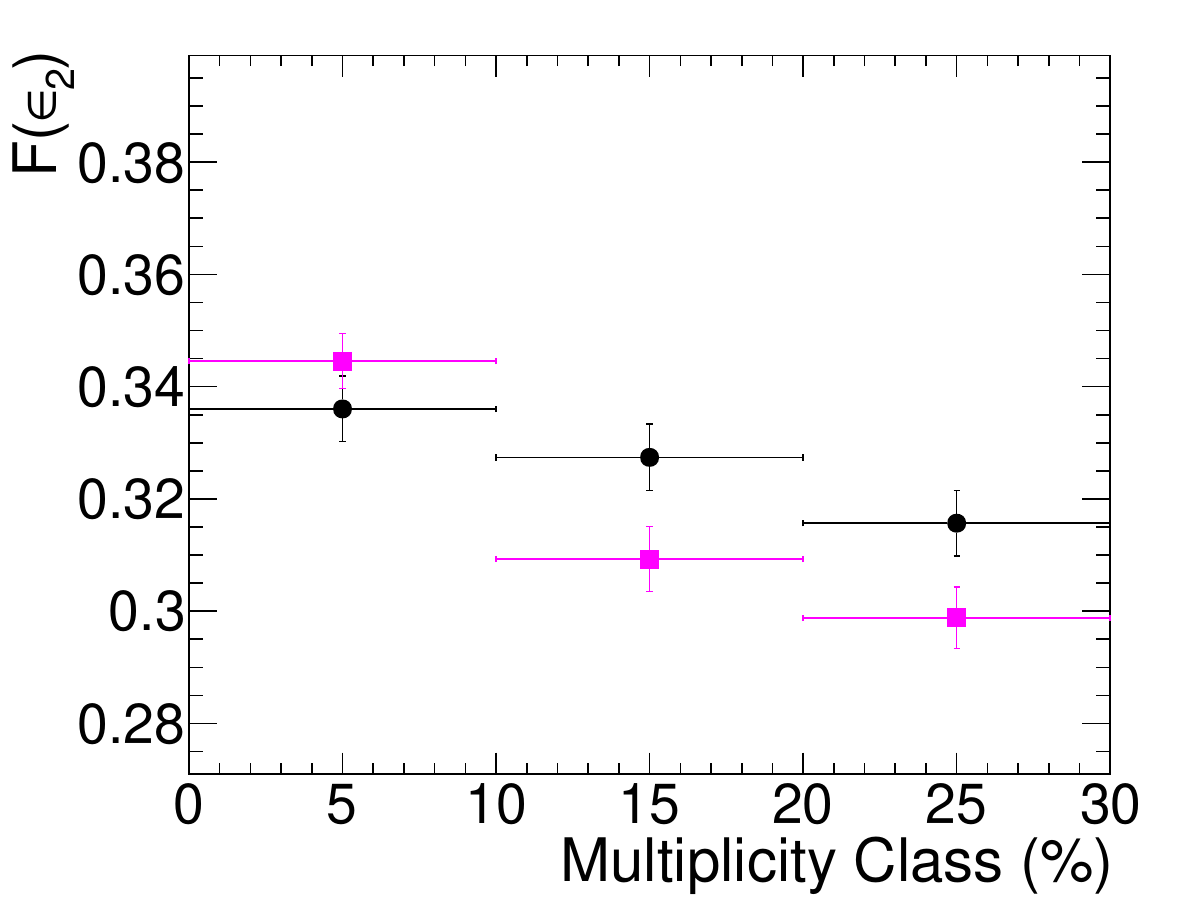}
    \includegraphics[scale=0.26]{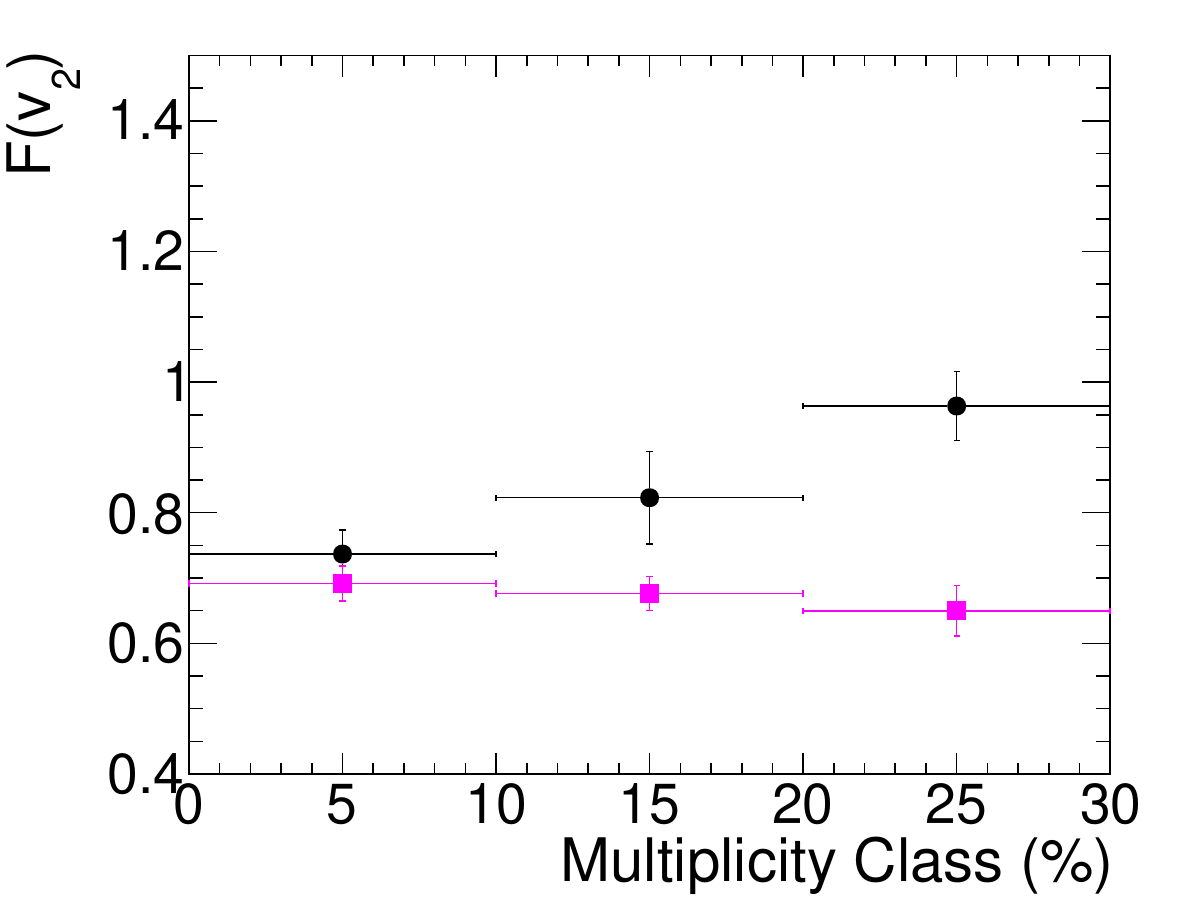}
    \caption{Upper panel shows the average value of eccentricity (left) and elliptic flow (right) in O-O collisions at $\sqrt{s_{\rm NN}}=7$ TeV using IPGlasma+MUSIC+iSS+UrQMD. The lower panel shows the corresponding relative fluctuations~\cite{Prasad:2024ahm}.}
    \label{fig:1}
\end{figure}
In this study, we use the IPGlasma+MUSIC+iSS+UrQMD model to generate O-O collisions at $\sqrt{s_{\rm NN}}=7$ TeV for both Woods\,--\,Saxon and $\alpha$-clustered nuclear profile. For the estimation of elliptic flow fluctuation, we employ the multi-particle Q-cumulant method~\cite{Prasad:2024ahm, Bilandzic:2010jr}. Model parameter details, along with different tunes and a detailed discussion of event and track selection cuts can be found in this earlier work, as well.

\section{Results and discussions}

Our focus is on to identify differences between nuclear structure models on the double-magic oxygen nucleus. Figure~\ref{fig:1} shows eccentricity, $\langle \epsilon_2\rangle$ and elliptic flow, $\langle v_2\rangle$ (upper panel) along with corresponding $F(\epsilon_2)$ and $F(v_2)$ (lower panel) in O-O collisions at $\sqrt{s_{\rm NN}}=7$~TeV. The values are compared between Woods\,--\,Saxon and $\alpha$-clustered nuclear structures of \textsuperscript{16}O nuclei. One observes higher values of $\langle \epsilon_2\rangle$ and $\langle v_2\rangle$ for the $\alpha$-clustered case as compared to Woods\,--\,Saxon nuclear profile. Further, as one moves from 0 to 30\% multiplicity class, $\langle \epsilon_2\rangle$ increases for both the nuclear profiles, while $\langle v_2\rangle$ shows a slightly decreasing trend. This could be attributed to the lower lifetime of the medium towards lower multiplicity events, which restricts the complete transformation of $\langle \epsilon_2\rangle$ to $\langle v_2\rangle$. 
$F(\epsilon_2)$ decreases from 0 to 30\% multiplicity class for both the profiles, while $F(v_2)$ shows an increasing feature for the Woods\,--\,Saxon case. For (10-30)\% multiplicity class, $F(\epsilon_2)$ and $F(v_2)$ are higher for the Woods\,--\,Saxon case. Interestingly, with a change in the nuclear density profile of the colliding \textsuperscript{16}O, both initial and final state second-order fluctuations change. This makes elliptic flow fluctuations one of the observables to study clustered density profiles in experiments.

\section{Summary}
We have studied geometrical and hydrodynamical evolution of small colliding system. Eccentricity, elliptic flow and their relative fluctuations in O-O collisions at $\sqrt{s_{\rm NN}}=7$~TeV were investigated using IPGlasma+MUSIC+iSS+ UrQMD model. We observe that both initial and final state anisotropies are sensitive to the choice of nuclear profile. We conclude that elliptic flow fluctuations are one of the sensitive observables to study the nuclear profile dependence in oxygen-oxygen collisions at the LHC.

\section*{Acknowledgment}
S.P., N.M., and R.S. sincerely acknowledge the DAE-DST, Government of India funding under Project No. SR/MF/PS-02/2021-IITI (E-37123). G. G. B. gratefully acknowledges the Hungarian NRDIO, 2021-4.1.2-NEMZ\_KI-2024-00058, 2024- 1.2.5-T\'ET-2024-00022 and Wigner Scientific Computing Laboratory.

\end{document}